\documentclass[a4paper]{jpconf}
\usepackage{graphicx}
\usepackage{subfigure}
\usepackage{wrapfig}

\begin{document}
\title{Impact of gluon damping on heavy-quark quenching}

\author{M.~Nahrgang$^{(1,2)}$, M.~Bluhm$^{(3)}$, P.~B.~Gossiaux$^{(1)}$, J.~Aichelin$^{(1)}$}

\address{(1) SUBATECH, Universit\'e de Nantes, EMN, IN2P3/CNRS, 4 rue Alfred Kastler, 44307 Nantes cedex 3, France\\
         (2) Frankfurt Institute for Advanced Studies (FIAS), Ruth-Moufang-Str. 1, 60438 Frankfurt am Main, Germany\\
         (3) Dipartimento di Fisica, Universit\`{a} degli Studi di Torino and INFN, Sezione di Torino, via Giuria 1, 10125 Torino, Italy}

\ead{nahrgang@subatech.in2p3.fr}

\begin{abstract}
In this conference contribution, we discuss the influence of gluon-bremsstrahlung damping in hot, absorptive QCD matter on the heavy-quark radiation spectra. Within our Monte-Carlo implementation for the description of the heavy-quark in-medium propagation we demonstrate that as a consequence of gluon damping the quenching of heavy quarks becomes significantly affected at higher transverse momenta.
\end{abstract}

\section{Introduction}

The fundamental properties of strongly interacting matter are investigated in ultra-relativistic heavy-ion collisions. With these experiments one aims at  understanding the basic processes in QCD matter at finite temperature $T$ and/or baryon density. For this purpose heavy quarks are a particularly clean probe of the formed matter because they are predominantly produced from initial hard scatterings of participating nucleons.  Subsequently they undergo collisional and radiative processes through interactions with the locally thermalized light quarks and gluons in the surrounding medium. 

These interactions with the medium constituents lead to a thermalization of heavy quarks with small transverse momentum $p_T$, while those with a large $p_T$ suffer from a significant in-medium energy loss. This is reflected in sensitive observables such as the nuclear modification factor $R_{\rm AA}$. Results from RHIC for the heavy-flavour decay non-photonic single electron $R_{\rm AA}$ as well as the $R_{\rm AA}$ of D-mesons~\cite{RHIC} indicate indeed a substantial quenching of heavy quarks at higher $p_T$ up to $10$~GeV. The data from the latest measurements at the LHC~\cite{Alice} was able to extend this $p_T$-range up to $30$~GeV for very central collisions. This offers a new and unique opportunity for testing our understanding of in-medium energy loss mechanisms, with the prospect of disentangling eventually the different contributions from collisional and radiative energy loss processes. 

Radiative processes are considered to be the dominant contribution to parton energy loss at larger parton energies, but they may become affected by coherence effects in the medium, cf.~the review in~\cite{Peigne09}. As a consequence of the destructive interference of radiation amplitudes from successive scatterings off medium constituents within the formation length $l_f$ of gluon-bremsstrahlung, the radiation spectrum is suppressed  for larger gluon energies $\omega$. This is the QCD-analog of the Landau-Pomerantschuk-Migdal (LPM) effect~\cite{Baier97}.  Other effects may alter the radiation pattern as well. For example, the dielectric polarization of the medium modifies the dispersion relation of radiated gluons which aquire an effective in-medium mass $m_g$. This leads to a significant reduction of the radiation spectrum in the soft-$\omega$ region~\cite{Kampfer00,Djordjevic03}. Similarly, dissipative effects in the QCD medium may diminish the radiative energy loss of heavy quarks as 
advocated in~\cite{Bluhm12}. 

In these proceedings, we discuss the dissipative effect of gluon damping on the radiation spectra of heavy quarks. Focussing on the implementation of the novel effect into our Monte-Carlo approach to the heavy-quark in-medium propagation~\cite{Gossiaux08,Gossiaux09} we show that damping mechanisms influence the $R_{\rm AA}$ of heavy quarks in a pronounced way.

\section{Gluon damping phenomenology}

In an absorptive plasma, mechanisms, which lead to the damping of bremsstrahlung gluons, can reduce the radiative energy loss of traversing, highly-energetic charges. This phenomenon was first demonstrated in~\cite{Bluhm11} for asymptotic electric charges in electro-magnetic plasmas by modelling the dispersive medium via a complex index of refraction and then advocated in~\cite{Bluhm12} to be of relevance also in the absorptive Quark-Gluon Plasma (QGP). In fact, dissipative effects such as quark--anti-quark pair creation or secondary gluon-bremsstrahlung generation introduce an additional scale in the medium. If this damping length scale $l_d$, which is related to the gluon damping rate $\Gamma$ as $l_d\simeq 1/\Gamma$, is smaller than or of the order of $l_f$ of the nascent gluon, absorptive mechanisms will reduce the probability for the formation of gluon-bremsstrahlung and, thus, influence the radiative energy loss of partons. This might be particularly important in the case of the large formation lengths in coherent emission processes. 

Even though the aforementioned mechanisms represent higher-order effects in perturbative QCD, they might be relevant for the matter investigated in the laboratory. For example, the damping rate due to secondary gluon-bremsstrahlung formation  is parametrically given as $\Gamma\sim g^4 T$ up to logarithmic corrections of order $\mathcal{O}(1/g)$~\cite{Bluhm12}. Consequently, higher temperatures imply a larger damping rate and thus a smaller $l_d$. Moreover, the formation length of gluon radiation~\cite{Bluhm12} increases for a given $x=\omega/E$ with increasing energy $E$ of the emitting parton. 
Thus, for large E (as long as $\Gamma>1/L$, where $L$ is the parton path length in the medium) and/or large $T$ dissipative effects will be important for the radiation spectra of partons, a situation that is likely to be encountered experimentally at the LHC. 

\section{Medium-modification of the heavy-quark radiation spectrum}

\begin{figure}[t]
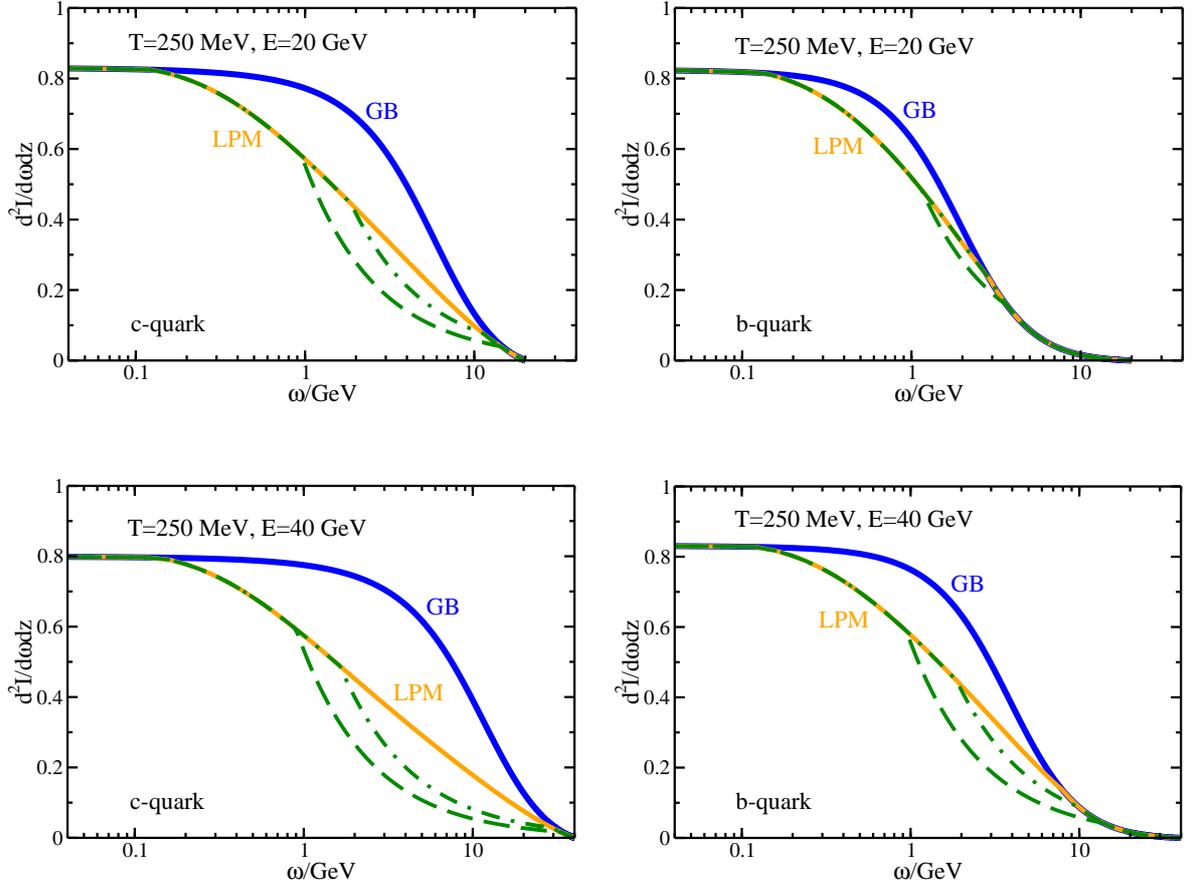

 \centering
 \subfigure{\includegraphics[width=0.47\textwidth]{SpCharm20.eps}} \hspace{2mm}
 \subfigure{\includegraphics[width=0.47\textwidth]{SpBeauty20.eps}}\\ \vspace{6mm}
 \subfigure{\includegraphics[width=0.47\textwidth]{SpCharm40.eps}} \hspace{2mm}
 \subfigure{\includegraphics[width=0.47\textwidth]{SpBeauty40.eps}}
 \caption{(Colour online) Suppression of the heavy-quark Gunion-Bertsch (GB) radiation spectrum (thick solid curves) due to LPM-effect (thinner solid curves) and gluon damping effect (dash-dotted curves for a damping rate $\Gamma=0.5\,T$ and dashed curves for $\Gamma=0.75\,T$, with the fixed gluon mass of $m_g=2\,T$). Left panels are for charm quarks with $m_c=1.5$~GeV, right panels for bottom quarks with $m_b=5.1$~GeV, while the upper (lower) row shows the results for a quark energy $E$ of $20$~GeV ($40$~GeV).}
 \label{fig:spectra}
\end{figure}
The Gunion-Bertsch (GB) gluon radiation spectrum~\cite{Gunion82} originating from single, independent scatterings of massless partons was generalized within scalar QCD to the case of massive partons in~\cite{Gossiaux10}. Medium-modifications of this heavy-quark GB-spectrum due to coherence effects were discussed in detail in~\cite{Gossiaux12a}. In Fig.~\ref{fig:spectra}, we show the radiation spectra of both cases for charm and bottom quarks at different parton energies $E$. 

In order to highlight quantitatively the additional impact of gluon damping effects on the bremsstrahlung spectrum off heavy quarks, we make use of a scaling ansatz advocated in~\cite{Galitsky64} according to which the global radiation intensity becomes modified due to medium effects via $d^2I=d^2I_{GB}\cdot \tilde{l}_f/l_f^0$. 
Here, the scale $l_f^0\simeq 2x(1-x)E/(m_g^2+x^2m_s^2)$ with the parton mass $m_s$ denotes the formation length in a single, independent scattering process \cite{Bluhm12,Gossiaux12a} and $\tilde{l}_f=\min(l_d,l_f)$ is given by the minimum between the damping length $l_d$ and the in-medium formation length $l_f$ discussed in~\cite{Bluhm12} fulfilling $l_f\leq l_f^0$.
 Figure~\ref{fig:spectra} exhibits the correspondingly modified spectra for two different values of $\Gamma$. As evident, damping effects may reduce the heavy-quark radiation spectrum significantly in an intermediate $\omega$-region, where the influence of the dissipative effects increases with both increasing $\Gamma$ and $E$. This reduction is stronger than the reduction due to the LPM-effect. Moreover, with larger $\Gamma$ and larger $E$ the spectra become more and more quark-mass independent.

\section{Influence of gluon damping on the quenching of heavy quarks}

\begin{wrapfigure}[19]{r}[0pt]{0.55\textwidth}
 \centering
  \includegraphics[width=0.55\textwidth]{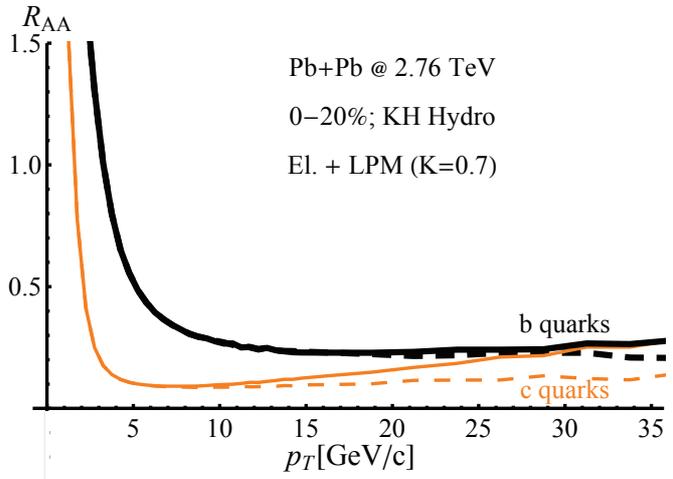} 
\caption{(Colour online) $R_{\rm AA}$ of charm (thin curves) and bottom quarks (thick curves) with (solid) and without (dashed) the effect of gluon damping. The damping rate is $\Gamma=0.75\,T$. For details cf.~\cite{Gossiaux12b}.}
  \label{fig:raa}
\end{wrapfigure}

To study the consequences of gluon damping on observables, we include the modified radiation spectra illustrated above in the Monte-Carlo approach~\cite{Gossiaux08,Gossiaux09} to the heavy-quark propagation in the medium. The interaction of the heavy quarks with the light partons in the QGP amounts in collisional and radiative processes, which are described in detail in~\cite{Gossiaux08,Gossiaux09,Gossiaux10,Gossiaux12a}, while the evolution of the medium is obtained from the fluid dynamic expansion presented in~\cite{KolbHeinz}. Any theoretical uncertainty affecting the model is assumed to be captured in one global factor K which rescales the interaction rates and which is obtained from calibration to the available RHIC data. In this spirit it is possible to reproduce the available heavy-flavour data at RHIC energies for both purely collisional and collisional plus radiative (without damping) energy loss scenarios as was demonstrated in~\cite{Gossiaux08,Gossiaux10,Gossiaux12a}. At LHC energies, the $p_T$-range is extended and we are able to identify different trends in the $R_{\rm AA}$ for these two scenarios, cf.~\cite{Gossiaux12b}. 
Counter-intuitively, available D-meson data from central $Pb+Pb$ collisions at the LHC~\cite{Alice} seem to favour the purely collisional energy loss scenario~\cite{Gossiaux12b}. This indicates that a reduction of the radiative energy loss component would be necessary in order to reconcile the model with the data. As illustrated in Fig.~\ref{fig:raa}, one possibility for such a reduction could be the inclusion of gluon damping mechanisms. 
Here, we show the $R_{\rm AA}$ of charm and bottom quarks for the collisional plus radiative energy loss scenario (dashed curves), which includes the LPM effect but no dissipative effects. The inclusion of gluon damping mechanisms results in the solid curves. As evident, when one neglects gluon damping effects, the $R_{\rm AA}$ stays almost flat for $p_T\geq 5$~GeV for the charm quarks and for $p_T\geq 15$~GeV for the heavier bottom quarks. Once gluon damping is taken into account, the nuclear modification factor increases visibly at higher transverse momenta. In the shown $p_T$-range predominantly charm quarks are affected, while the effect sets in for bottom quarks only at larger $p_T$. Details, however, depend on the value of the gluon damping rate.

\section{Conclusions}
We discussed the influence of gluon damping in the absorptive QGP on the quenching of heavy quarks in ultra-relativistic heavy-ion collisions. In addition to the known modification of radiation spectra due to coherence effects, gluon damping effects reduce significantly the spectra in an intermediate $\omega$-region for large gluon damping rates $\Gamma$ and/or large parton energies $E$. Consequently, in the absorptive QGP heavy quarks become less quenched at large transverse momenta. 
 With increasing $p_T$, the quenching seems to become quark-mass independent which would be visible in a prominent behaviour of the heavy-flavour meson $R_{\rm AA}$ as advocated in~\cite{Gossiaux12b}.
 \vspace{0.1cm}

MN and MB thank the organisers of the conference {\it Heavy Ion Collisions in the LHC Era, Quy Nhon, Vietnam, 16-20 July 2012}, for financing their participation. MB acknowledges the financial support received from the SaporeGravis Network of Hadron Physics I3 (I3HP3). MN thanks her parents for financing her flight to Vietnam.

\end{document}